\documentclass[11pt]{article}

\usepackage[dvips]{graphicx}
\usepackage{amsfonts}
\usepackage{amssymb}
\usepackage{subeqn}
\usepackage{amsbsy}

\newcommand{\R}{{\mathbb R}}
\newcommand{\C}{{\mathbb C}}

\newcommand{\cF}{{\cal F}}

\newcommand{\beq}{\begin{equation}}
\newcommand{\eeq}{\end{equation}}

\newcommand{\pa}{\partial}

\newcommand{\bx}{{\bf x}}

\newcommand{\Imag}{\mathop{\mathrm{Im}}}
\newcommand{\Sg}{\mathop{\mathrm{sgn}}}
\newcommand{\Cs}{\mathop{\mathrm{const.}}}
\newcommand{\Ai}{{\mathop{\mathrm{Ai}}}}
\newcommand{\Bi}{{\mathop{\mathrm{Bi}}}}
\newcommand{\ph}{{\mathop{\mathrm{ph\,}}}}

\newcommand{\smfr}[2]{{\textstyle{#1 \over #2}}}

\newcommand{\rmi}{\mathrm{i}}
\newcommand{\rmd}{\mathrm{d}}

\begin{document}

\title{The Evanescent Waves in Geometrical Optics and the Mixed Hyperbolic--Elliptic Type Systems}

\author{
Enrico De Micheli \\
\normalsize IBF -- Consiglio Nazionale delle Ricerche \\
\normalsize Via De Marini, 6 - 16149 Genova, Italy. \\
e-mail: demicheli@ge.cnr.it
\and
\\ Giovanni Alberto Viano \\
\normalsize Dipartimento di Fisica - Universit\`a di Genova \\
\normalsize Istituto Nazionale di Fisica Nucleare - sez. di Genova \\
\normalsize Via Dodecaneso, 33 - 16146 Genova, Italy. \\
e-mail: viano@ge.infn.it}

\date{}
\maketitle

\begin{abstract}
In this article we describe the generation of the
evanescent waves which are present in the rarer medium at total
reflection by using a mixed--type system, the Ludwig system, which
leads naturally to consider a complex-valued phase. The Ludwig
system is derived from the Helmholtz equation by using an
appropriate modification of the stationary phase procedure: the
Chester, Friedman and Ursell's method. The passage from the
illuminated to the shadow region is described by means of the ray
switching mechanism based on the {\it Stokes phenomenon} applied
to the Airy function. Finally, the transport system connected to
the Ludwig eikonal system is studied in the case of linear
wavefronts and the existence of the Goos--H\"anchen effect is
proved.
\end{abstract}


\section{Introduction}
\label{se:introduction}
The study of evanescent waves attracts nowadays the attention of many authors in view of their great relevance
in several physical and technological problems \cite{Scott,Sides,Temple}.
In this article we want to reconsider what is probably
the first and more elementary process, which generates evanescent waves: the total reflection
at the interface between an optically denser and a rarer medium.
In spite of the fact that this phenomenon has been deeply
studied, starting from Newton \cite{Newton}, still some open theoretical questions remain. These ambiguities
emerge with some evidence from what Sommerfeld writes, about this subject, in his classical textbook \cite{Sommerfeld}.
Concerning the propagation along the boundary, he writes: {\it ``An actual energy flow takes place parallel to the
boundary surface. This seems to contradict both the name ``total reflection'' and our oft--repeated statement
that no energy is lost in the process. We must, however, consider the fact that we have always
performed our calculations for the ideal case of an infinitely wide wave front. For the actual, laterally
restricted waves, energy can very well pass from the denser to the rarer medium or, respectively flow back
from the rarer into the denser medium at the lateral boundaries of the wave. This is the energy which is
transported parallel to the boundary surface or, so to say, meanders about it''}. Next, he quotes the experiments of
Goos and H\"anchen \cite{Goos}, which show the penetration of the incident ray into the rarer medium.

At this point two questions arise:
\begin{itemize}
\item[(i)]
Is it possible to find a system which describes globally the propagation of rays in the various domains: in the
illuminated region (denser medium), in the shadow (rarer medium) and also along the boundary?
\item[(ii)] Can this global representation of ray propagation explain the penetration of the incident wave into
the rarer medium without introducing complex--valued transmission angles in a formal way?
\end{itemize}

We shall try to answer the previous questions limiting ourselves to the scalar representation of the light
(which corresponds to linearly TE polarized light).
In what follows we will describe phenomena which are related to the generation of evanescent waves
at total reflection, as the plane wave propagation along the interface and the Goos--H\"anchen effect.
Of course, this does not exhaust the whole phenomenology; other effects and waves which can be present, e.g.,
surface waves excited at the interface \cite{Tew} and waves exhibiting unusual properties (P$^*$ wave) \cite{Babich1},
will not considered here.

It is well--known that geometrical optics is given by the leading term of an asymptotic expansion of a solution
of the wave equation, which is usually derived by using the stationary phase method. One then introduces the
concept of ray, which is described by the so--called eikonal equation.
In the present article we aim at showing that the propagation along the boundary, which is a peculiar feature of
total reflection, cannot be treated within the framework of the standard geometrical optics, but by the use of
an extended formulation of the latter based on the Ludwig eikonal system. Indeed, as we shall show in
Section \ref{se:complex}, the standard stationary phase method cannot be used to describe
the ray propagation along the boundary, when the incident ray reaches
the critical angle, since two stationary points coalesce at the interface between the two media.
One is then forced to use
a modification suggested by Chester, Friedman and Ursell (CFU method) \cite{Chester}, which leads to a non--standard
eikonal equation, which was first introduced by Ludwig \cite{Ludwig} in a quite general context.

The paper is organized as follows: In Section \ref{se:collision}
we start from the Helmholtz equation and derive the Ludwig system
by means of the CFU method. In Section \ref{se:complex}
we discuss in detail the mixed--type (hyperbolic--elliptic) character
of the Ludwig eikonal system. In Section \ref{se:stokes} the Stokes switching mechanism, which allows us to pass from
the two--ray illuminated region to the one--ray shadow region, comes into play.
In Section \ref{se:generation} we study in detail
the propagation of the waves in the various domains, in the case of linear wavefronts.
We show that the Ludwig eikonal system can, indeed, furnish a global
description of the phenomena and explain the generation of the evanescent waves in the shadow region.
Moreover, from the analysis of the Ludwig transport system we describe the Goos--H\"anchen lateral shift
of a reflected ray at total reflection \cite{Goos}, which is related to the penetration of the incident ray
into the rarer medium.
In conclusion, we can give a first, partial positive answer to the
questions posed earlier in the limit of the scalar treatment of the problem. Finally, in the Appendix we
illustrate in detail how to derive the Stokes switching mechanism by means of the Borel summation method.

\section{Collision of stationary points and the Ludwig eikonal system}
\label{se:collision}
Let us start from the Helmholtz equation, which reads
\beq
\label{uno}
\nabla^2 \psi + k^2 n^2 \psi = 0,
\eeq
where $k$ is the free--space wavenumber, and $n$ is the refractive index of a homogeneous medium.
We look for a solution to (\ref{uno}) of the following form:
\beq
\label{due}
\psi(\bx,k)=\int A(\bx,\beta)\,e^{\rmi k\Phi(\bx,\beta)}\,\rmd\beta,
\eeq
$(\bx\in\R^3)$. The principal contribution to $\psi(\bx,k)$ as $k\rightarrow\infty$, corresponds to the
stationary point of $\Phi$, in the neighborhood of which the exponential $\exp(\rmi k\Phi)$ ceases to
oscillate rapidly. These stationary points can be obtained from the equation
$\pa\Phi(\bx,\beta)/\pa\beta = 0$ (provided that $\pa^2\Phi(\bx,\beta)/\pa\beta^2 \neq 0$).
Let us suppose for the moment that at each point $\bx$
we have only one stationary point of $\Phi$, then the asymptotic expansion of $\psi$ as $k\rightarrow\infty$
can be written as follows:
\beq
\label{tre}
\psi(\bx,k)\simeq e^{\rmi k\Phi(\bx,\beta_0)}\sum_{m=0}^\infty\frac{A_m}{(\rmi k)^m},
\eeq
where $\beta_0$ is the unique stationary point of $\Phi$. The leading term of expansion (\ref{tre}) reads
\beq
\label{quattro}
\psi(\bx,k) \simeq A_0(\bx,\beta_0)\,e^{\rmi k\Phi(\bx,\beta_0)},
\eeq
where
\beq
\label{cinque}
A_0(\bx,\beta_0) = A(\bx,\beta_0)\left(\left|\frac{\pa^2\Phi}{\pa\beta^2}\right|^{-1/2}\right)_{\beta=\beta_0}
\exp\left[\rmi\frac{\pi}{4}\Sg(\pa^2\Phi/\pa\beta^2)_{\beta=\beta_0}\right].
\eeq
Substituting the leading term (that hereafter will be written as $A\exp(\rmi k\Phi)$, omitting the
subscript zero), into (\ref{uno}), collecting powers of $(\rmi k)$, and then equating to zero their
coefficients, we obtain two equations: the eikonal (or Hamilton--Jacobi) equation:
\beq
\label{sei}
\left(\nabla\Phi\right)^2 = n^2,
\eeq
and the transport equation
\beq
\label{sette}
\nabla \cdot (A^2\nabla\Phi) = 0,
\eeq
whose physical meaning is that the current density is conserved.
The surface $\Phi=\Cs$ is called {\it constant--phase} surface.

Now, suppose that two stationary points coalesce.
In such a case the standard asymptotic method and the corresponding asymptotic expansion fail, and
a different strategy must be adopted. The procedure suggested by Chester, Friedman and Ursell \cite{Chester}
consists in driving the phase function $\Phi$ into a new more convenient form by a suitable change of the
integration variable $\beta\leftrightarrow\xi$ (provided $\Phi_{\beta\beta\beta}\neq 0$):
\beq
\label{otto}
\Phi(\bx,\beta) = u(\bx) + f(\bx,\xi).
\eeq
After this change of variable, integral (\ref{due}) reads:
\beq
\label{nove}
\psi(\bx,k)= e^{\rmi ku(\bx)} \int g(\bx,\xi)\,e^{\rmi kf(\bx,\xi)}\,\rmd\xi.
\eeq
This expression is very similar to integral (\ref{due}) with the phase function $\Phi$ replaced
by the function $f$, and with an additional oscillatory factor $\exp(\rmi ku)$ in front.

Now, suppose that $\pa\Phi/\pa\beta$ vanishes at two distinct stationary points $\beta_1(\bx)$ and $\beta_2(\bx)$.
We want to choose a transformation such that to these points there correspond points where $\pa f/\pa\xi$
vanishes, which are symmetric with respect to $\xi=0$. The result can be achieved by setting
\beq
\label{dieci}
f(\bx,\xi) = -\smfr{1}{3} \xi^3 + v(\bx)\,\xi.
\eeq
In fact, $\pa f/\pa\xi=-\xi^2+v(\bx)$, and the equation $\pa f/\pa\xi=0$ yields $\xi=\pm\sqrt{v}$. Then from
equalities (\ref{otto}) and (\ref{dieci}) (setting $\xi=\sqrt{v}$ and $\xi=-\sqrt{v}$), we obtain the
following relationships:
\begin{subequations}
\label{undici}
\begin{eqnarray}
\Phi(\bx, \beta_1) & = & u(\bx) + \smfr{2}{3}v^{3/2}(\bx), \label{undicia} \\
\Phi(\bx, \beta_2) & = & u(\bx) - \smfr{2}{3}v^{3/2}(\bx), \label{undicib}
\end{eqnarray}
\end{subequations}
that yield
\begin{subequations}
\label{dodici}
\begin{eqnarray}
u(\bx) & = & \smfr{1}{2}\left[\Phi(\bx,\beta_1)+\Phi(\bx,\beta_2)\right], \label{dodicia} \\
\smfr{2}{3}v^{3/2}(\bx) & = & \smfr{1}{2}\left[\Phi(\bx,\beta_1)-\Phi(\bx,\beta_2)\right]. \label{dodicib}
\end{eqnarray}
\end{subequations}
In the case $\beta_1=\beta_2$, we have $v(\bx)=0$ and $u(\bx)=\Phi(\bx,\beta_1)=\Phi(\bx,\beta_2)$. If
equalities (\ref{dodici}) are satisfied, then the transformation $\xi\leftrightarrow\beta$
is uniformly regular and 1--1 near $\xi=0$ (see \cite{Chester}).
From (\ref{nove}) and (\ref{dieci}) it follows that the leading terms in the expression of
$\psi(\bx,k)$, for large $k$, can be written in terms of the Airy function $\Ai(\cdot)$ and of its derivative
$\Ai'(\cdot)$. We have
\beq
\label{tredici}
\psi(\bx,k) \simeq e^{\rmi ku(\bx)}\left[g_0(\bx)\int e^{\rmi k(v\xi-\xi^3/3)}\,\rmd\xi+
h_0(\bx)\int\xi e^{\rmi k(v\xi-\xi^3/3)}\,\rmd\xi\right],
\eeq
where $g_0(\bx)$ and $h_0(\bx)$ are respectively the first terms of the two formal asymptotic series:
$\sum_{m=0}^\infty g_m(\bx)/k^m$, and $\sum_{m=0}^\infty h_m(\bx)/k^m$. Then, by using the integral representation
of the Airy functions we finally obtain:
\beq
\label{quattordici}
\psi(\bx,k) \simeq 2\pi e^{\rmi ku(\bx)}\left[\frac{g_0(\bx)}{k^{1/3}} \Ai(-k^{2/3}v) +
\frac{h_0(\bx)}{\rmi k^{2/3}} \Ai'(-k^{2/3}v)\right],
\eeq
which, substituted into (\ref{uno}), yields a set of four equations.
In what follows we will focus only on those equations which correspond to the eikonal
equation (\ref{sei}), which generate, in a general geometrical context,
a complex--valued phase and, consequently, evanescent waves.
The equations which correspond to the transport equation (\ref{sette}) will be analyzed in Section
\ref{se:generation}, in the particular case of the propagation of linear wavefronts.

The system corresponding to the eikonal equation reads:
\begin{subequations}
\label{quindici}
\begin{eqnarray}
\left| \nabla u\right|^2 + v \left| \nabla v\right|^2 &=& n^2, \label{quindicia} \\
\nabla u \cdot \nabla v &=& 0, \label{quindicib}
\end{eqnarray}
\end{subequations}
which is a fully nonlinear, first order partial differential system ruling $u$ and $v$.
Hereafter system (\ref{quindici}) will be referred to as the {\it Ludwig eikonal system} \cite{Ludwig}.

\section{Complex rays and mixed--type systems}
\label{se:complex}
The Ludwig system has been derived in connection with the problem of obtaining a uniform asymptotic
expansion at a caustic \cite{Ludwig}.
Let us recall that the caustic is the envelope of a family of rays \cite{Stavroudis}: i.e., (i)
for every point on the caustic it is possible to give a curve of the family which is tangent to the
caustic on this point; (ii) no curve of the family has a segment in common with the caustic \cite{Pogorelov}.
When total reflection occurs, even at the critical angle, at the
boundary between the media we do not have a caustic in the strict sense of the definition given above.
Nevertheless, beyond the critical angle, we have propagation along the boundary,
and the constant phase line of this ray is orthogonal to the boundary.
If the latter is the horizontal
axis, then the wavefront is vertical. It can be reached either by varying the constant phase lines orthogonal
to the reflected rays, from the limiting angle up to the grazing angle, or by varying, in a similar way,
the constant phase lines orthogonal to the direction of the incident rays. We can thus say that the
wavefront of the ray running along the boundary is the vertical tangent to the cusps generated by
the envelope of the wavefronts of the reflected rays when it touches, on the horizontal axis, the
envelope of the wavefronts of the incident rays. Therefore we can regard the boundary as the locus of these cusps.
We can thus say that on each point of the boundary two stationary points coalesce and the standard
stationary phase method cannot be used.
We are then led to the CFU method and, accordingly, to describe
the geometry of the rays by the use of the Ludwig eikonal
system. We can then work out the problem through the system (\ref{quindici})
taking, for the refractive index $n$, a value $n>1$ in the denser medium and, for the sake of simplicity
and without loss of generality, the value $n=1$ in the rarer one. At the boundary there is a discontinuity
in $n$.

We want now to study the mixed character of the Ludwig system by working out the problem in $\R^2$ and
using Cartesian orthogonal axes, whose coordinates will be denoted by $x$ and $y$.
First we write the characteristic determinant (see also \cite{Magnanini1,Magnanini3}):
\beq
\label{sedici}
D = \left |
\begin{array}{cccc}
2u_x & 2u_y & 2vv_x & 2vv_y \\
\rmd x   & \rmd y   & 0     & 0     \\
v_x  & v_y  & u_x   & u_y   \\
0    & 0    & \rmd x    & \rmd y
\end{array}
\right | =
2\left[(u_x \rmd y - u_y \rmd x)^2 - v (v_x \rmd y - v_y \rmd x)^2\right].
\eeq
Then, the equation of the characteristics for the Ludwig system reads
\beq
\label{n5}
(u_y^2-vv_y^2)\left(\frac{\rmd x}{\rmd y}\right)^2+2(vv_xv_y-u_xu_y)\left(\frac{\rmd x}{\rmd y}\right)+(u_x^2-vv_x^2)=0,
\eeq
whose discriminant $\Delta$ is
\beq
\label{n6}
\Delta = (vv_xv_y-u_xu_y)^2-(u_y^2-vv_y^2)(u_x^2-vv_x^2)=v(u_yv_x-u_xv_y)^2=vJ^2,
\eeq
where $J$ is the Jacobian determinant of the transformation $(x,y) \leftrightarrow (u,v)$.
Provided $J \neq 0$, and according to the sign of $\Delta$, we can distinguish among the following cases:
For $v>0$, i.e., $\Delta>0$, we are in the hyperbolic case (the characteristics are real and distinct);
for $v=0$, i.e., $\Delta=0$, we are in the parabolic case (the characteristics are real and coincident);
for $v<0$, i.e., $\Delta<0$, we are in the elliptic case (there are no real characteristics).
We have a system of mixed type \cite{Bers}.

\begin{figure}[t]
\centering
\includegraphics{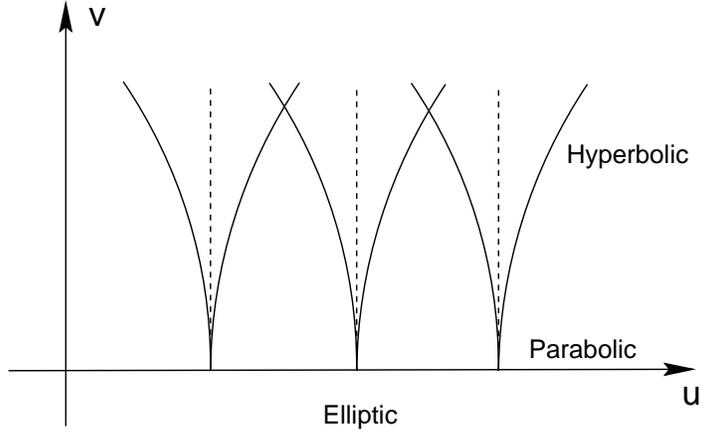}
\caption{\label{figure_1}\small The cubic curves (\ref{diciotto}) in the $(u,v)$--plane.}
\end{figure}

Assuming $v>0$ (corresponding to the illuminated region), Eq. (\ref{n5}) leads to the following characteristic curves:
\beq
\label{diciassette}
\rmd x:\rmd y = (u_x \pm \sqrt{v} v_x) : (u_y \pm \sqrt{v} v_y).
\eeq
Then, we can introduce, in the $(u,v)$--plane, the curves
\beq
\label{diciotto}
\Phi^{(\pm)}(u,v) = u \pm \smfr{2}{3} v^{3/2} = \Cs,
\eeq
which are branches of cubic curves with negative ($\Phi^{(+)}$) and positive ($\Phi^{(-)}$) slopes,
respectively (see Fig. \ref{figure_1}).
Let us note that at $v=0$, these curves have cusps with vertical tangent.
From (\ref{diciassette}) and (\ref{diciotto}) we have
\beq
\label{venti}
\frac{\rmd x}{\rmd y} = \frac{u_x \pm \sqrt{v} v_x}{u_y \pm \sqrt{v} v_y} = \frac{\Phi^{(\pm)}_x}{\Phi^{(\pm)}_y},
\eeq
and, furthermore, it is easy to verify that $|\nabla(u \pm \smfr{2}{3}v^{3/2})|^2=n^2$. Thus, it can be concluded that
$\nabla(u \pm \smfr{2}{3}v^{3/2})$ give the directions of the rays, which are orthogonal to the
curves $\Phi^{(\pm)}=\Cs$ (see Fig. \ref{figure_1}).
These are real rays, as real are the characteristic curves of the Ludwig system:
for $v>0$, this system is of {\it hyperbolic type}, and the lines $\Phi^{(\pm)}=\Cs$
are the constant phase curves in the $(u,v)$--plane.

In the case $v=0$ the system is of parabolic type; we are at the boundary between the two media
and the refractive index presents a discontinuity.
This situation can be still analyzed as a limiting case: the lines of constant phase are the vertical tangents to
the cusps generated by the curves $\Phi^{(\pm)}=\Cs$. The rays are real and orthogonal to the level
lines $u=\Cs$.

Finally, let us consider the case $v<0$, which corresponds to the shadow region and
in which the system is elliptic. From (\ref{diciotto}) we see that the phase $\Phi^{(\pm)}(u,v)$ becomes:
$\Phi^{(\pm)}(u,v)=u\mp\rmi\frac{2}{3}(-v)^{3/2}$: i.e., it becomes complex--valued.
Equation (\ref{quindicib}) informs us that the gradients of $u$ and $v$ are orthogonal; then, where
$J \neq 0$, this amounts to write
\beq
\label{ventidue}
\frac{1}{|\nabla u|}
\left[
\begin{array}{c}
u_x \\
u_y
\end{array}
\right]
=
\frac{1}{|\nabla v|}
\left[
\begin{array}{c}
v_y \\
-v_x
\end{array}
\right],
\eeq
from which the equalities
\begin{subequations}
\label{ventitre}
\begin{eqnarray}
u_x &=& \rho \, v_y, \label{ventitrea} \\
u_y &=& -\rho \, v_x, \label{ventitreb}
\end{eqnarray}
\end{subequations}
follow, with $\rho(x,y)=|\nabla u|/|\nabla v| = ((1/|\nabla v|^2)-v)^{1/2} > 0$
(recall that in the shadow region $n=1$). From (\ref{quindicia}) we have that inside
the shadow region $|\nabla u|^2 \geq 1$, the equality holding only when $|\nabla v| = 0$.
Then, it follows that $\rho$ is bounded away from zero,
provided that $|\nabla v|$ is bounded; in this case equations
(\ref{ventitre}) form a Beltrami system expressing the fact that
the mapping $(x,y)\rightarrow (u,v)$ is conformal with respect to
the Riemannian metric
\beq
\label{n4}
\rmd s^2 = \frac{1}{\rho} (\rmd x^2 + \rmd y^2),
\eeq
i.e., that under this mapping an angle $\alpha$ in
the $(x,y)$--plane measured by the metric (\ref{n4}) in taken into
an angle $\alpha$ in the $(u,v)$--plane, measured this time in the
usual way \cite{Bers}. Moreover, let us consider a compact
subdomain of the shadow region defined by $0 < (-v_0) \leqslant (-v)
\leqslant (-v_1) < +\infty$; there inside we have that $\rho$ is
bounded and bounded away from zero, provided $J \neq 0$ and
$|\nabla u|$ bounded: i.e., there exists a constant $Q$ such
that
\beq
\label{nnn1}
0 < \frac{1}{Q} \leqslant \frac{1}{\rho} + \rho \leqslant Q < +\infty.
\eeq
In this case system (\ref{ventitre}) is
uniformly elliptic, and the transformation $(x,y)\rightarrow
(u,v)$ maps infinitesimal circles into small ellipses of uniformly
bounded eccentricity: i.e., it is a quasi--conformal mapping
\cite{Bers}. Moreover, the following differential inequality is
satisfied:
\beq
\label{nnn2}
u_x^2+u_y^2+v_x^2+v_y^2 \leqslant \left (\frac{1}{Q}+Q \right ) J,
\eeq
where $J$ is the Jacobian of the
transformation associated to equations (\ref{ventitre}). Notice
that since $|\nabla v|^2 = (|\nabla u|^2 -1) / (-v)$, the gradient
of $v$ tends to vanish as $v$ increases into the shadow region, so
that any oscillation is smoothed down by the elliptic character of
the Ludwig system for $v < 0$.

Summarizing, we have that for $v>0$ (hyperbolic case) the values of the phase are given by
$\Phi^{(\pm)}=u \pm \frac{2}{3}v^{3/2}$;
for $v=0$ (parabolic case) we simply have $\Phi=u$; for $v<0$ (elliptic case) the complex--valued phase
$\Phi^{(\pm)}=u\mp\rmi\frac{2}{3} (-v)^{3/2}$ naturally emerges; finally, departing from the boundary
between the two media, that is in any compact subdomain of the shadow region, the system becomes uniformly elliptic.

It is important to stress here that the complex--valued phase emerges as a direct consequence
of our formulation of the problem, which leads to consider the Ludwig system: its elliptic character
in the shadow generates a complex--valued phase.
This situation is deeply different from the approaches of complexification of the real--ray
theory, in which the complex values of the phase are considered from the
beginning and then, by considering all the ray intersections with the real space, aim at selecting all the relevant
complex ray contributions to the solution of the physical problem
(see \cite{Chapman} for a detailed discussion of {\it pros and cons} of this kind of approach).

~

\noindent
{\em Remark}. We have just seen that by fixing $u$, then $v$ is determined through the system
(\ref{quindici}): we have a map $u\rightarrow v$. We can find the map $v\rightarrow u$ as well.
Both maps can be viewed as B\"{a}cklund--type transformations which couple (\ref{quindicia}) and
(\ref{quindicib}). For instance, the map $v\rightarrow u$, in the shadow region, can be expressed through the
following B\"{a}cklund--type transformation \cite{Magnanini1,Magnanini2,Rogers}:
\beq
\label{n9}
\nabla u = \mp \frac{\sqrt{1-v |\nabla v|^2}}{|\nabla v|}
\left(
\begin{array}{c}
-v_y \\
v_x
\end{array}
\right) ~~~~~~ (v < 0),
\eeq
which couples the two relationships:
\begin{itemize}
\item[(a)] orthogonality of $\nabla u$ with respect to $\nabla v$ (see (\ref{quindicib}));
\item[(b)] the equality: $|\nabla u|^2=1-v |\nabla v|^2$ (see (\ref{quindicia})).
\end{itemize}

\section{Stokes phenomenon and asymptotic behavior}
\label{se:stokes}
We can now draw our attention to the amplitude term in solution (\ref{quattordici}) (i.e., the term in brackets),
and in particular to the Airy function $\Ai(z)$, which satisfies the Airy equation
\beq
\label{ventinove}
\frac{\rmd^2w}{\rmd z^2}=zw.
\eeq
For large $|z|$, the solutions to (\ref{ventinove}) can be approximated by a linear combination of the functions
(see the appendix):
\beq
\label{trenta}
w_\pm^{(\rm a)} = z^{-1/4}\,\exp\,(\pm \smfr{2}{3} z^{3/2}).
\eeq

Evidently $w_+^{(\rm a)}(z)$ and $w_-^{(\rm a)}(z)$ are multivalued functions of the complex variable $z$
with a branchpoint
at $z=0$; instead, the solutions $w(z)$ of (\ref{ventinove}) are entire function of $z$
because the coefficient of $w(z)$ in (\ref{ventinove}), i.e., $z$, is entire.
Therefore, when we turn once around the point $z=0$, $w(z)$ will return to its original value,
but $w_+^{(\rm a)}$ and $w_-^{(\rm a)}$ will not. It follows that if a specific solution
$w(z)$ of (\ref{ventinove}) is approximated at $z\neq 0$, by a linear combination
$c_1 w_+^{(\rm a)}+c_2 w_-^{(\rm a)}$,
then it cannot be approximated by the same linear combination at $ze^{2\pi\rmi}$,
that is after one turn around $z=0$.
The concept of approximation here involved must be domain--dependent: different approximations {\it live}
in different angular sectors of the complex $z$-plane.
This is basically the Stokes phenomenon \cite{Berry,Dingle,Olver}:
constructing an uniform approximation leads to retain small exponentials though their numerical value
is dominated by that of large exponentials; across Stokes lines, the multiplier of the small exponential changes
smoothly but rapidly \cite{Berry,Berry2,McLeod},
and the aim is to establish the relation between multipliers in the
different parts of the complex domain.
These facts are the price to pay when one wishes to represent
entire functions by multivalued functions. At first sight, it could appear strange to represent entire
functions by multivalued functions, however it is worth noting that the representation by multivalued
functions is usually the only way in which the wave character of a solution can be made clearly explicit;
this is the main reason of the pervasiveness of the Stokes phenomenon in many different contexts.

\begin{figure} [t]
\centering
\includegraphics[width=4.6in]{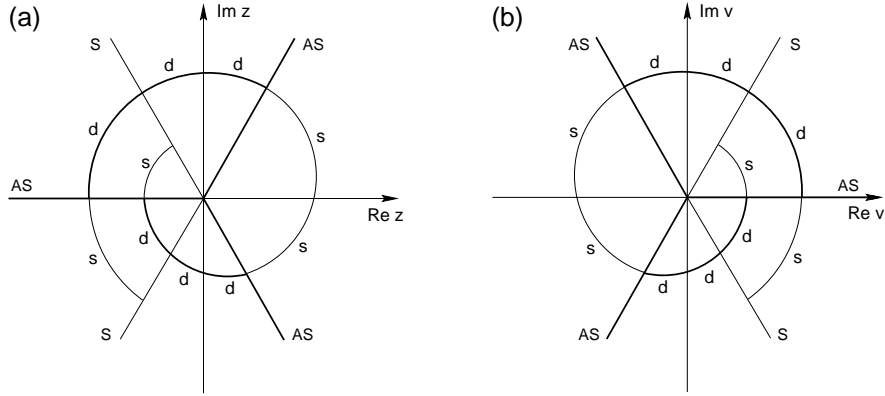}
\caption{\label{figure_2}\small (a): Representation of the asymptotic expansion of $\Ai(z)$.
Stokes lines (thin lines indicated by S); anti--Stokes lines (tick lines indicated by AS).
Dominant (tick lines labeled by `d') and subdominant (thin lines labeled by `s') rays.
(b): Representation of the asymptotic expansion of $\Ai(-k^{2/3}v)$
(see also \cite{Chapman}, Fig. 15 for an analogous picture).}
\end{figure}

In the case of the approximation of solutions to second order homogeneous differential equations,
and in particular of the Airy equation, which is our concern here,
the asymptotic expansion of a solution is, in general, a linear combination
of two complex exponentials of the type $\exp(\phi_\pm(z)), (z\in\C)$.
The usual prescription states that Stokes lines, i.e., the lines
where nonuniformity in the asymptotic expansion occur, are located where the phases of the two exponentials
are equal, i.e., when $\Imag [\phi_+(z)-\phi_-(z)] = 0$. However, not all the equal--phase lines are
Stokes lines and, in general, the problem is to decide which of these lines are actually Stokes lines.
In the Appendix, these and other questions related to the Stokes phenomenon are discussed extensively;
in particular, we show how to locate Stokes lines without
invoking the equal--phase argument above, and, accordingly, we construct the asymptotic
expansion of the function $\Ai(z)$ in the different angular domains, determining where and how ray--fields are
switched on and off (see formulae (\ref{A18}) and the related ray structure in Fig. \ref{figure_2}a).
The Stokes phenomenon plays a central role in our analysis in view of the fact that we must connect the
illuminated region, i.e., $v>0$, with the shadow, i.e., $v<0$; this connection can be achieved
by investigating what happens to the asymptotic approximation of the Airy function when we turn around the
point $v=0$, with $v$ regarded as a complex variable.
The Ludwig approximation (\ref{quattordici}) contains the Airy function $\Ai(-k^{2/3}v)$
(in this approximation we neglect the first derivative of the Airy function, and this is
legitimate in a region close to the boundary between the two media), whose ray structure is shown in Fig. \ref{figure_2}b.
At this point we apply the results shown with details in the Appendix and we note that for $v$ real and positive
we have two contributions of comparable amplitude, whereas for $v$ real and negative, i.e., in the shadow, we have
only one contribution with subdominant character. For $v \in \R^+$ we have
(see the definition of $w_\pm$ in (\ref{A10})):
\beq
\Ai(-k^{2/3}v)=\frac{1}{2\sqrt{\pi}}[w_-(-k^{2/3}v)+\rmi w_+(-k^{2/3}v)]~~~~~(v \in \R^+).
\label{illuminated}
\eeq
At high frequency (i.e., for $k\rightarrow+\infty$), we
can approximate $w_\pm$ with $w_\pm^{(\rm a)}$, obtaining for $v \in \R^+$:
\beq
\Ai(-k^{2/3}v)=\frac{1}{2\sqrt{\pi}}\frac{1}{k^{1/6}}e^{-\rmi\pi/4}
\left(\frac{e^{\rmi k\frac{2}{3}v^{3/2}}}{v^{1/4}}
+\frac{e^{\rmi\pi/2}\,e^{-\rmi k\frac{2}{3}v^{3/2}}}{v^{1/4}}\right)~~~~~(v \in \R^+).
\label{new-illuminated}
\eeq
Plugging the r.h.s. of formula (\ref{new-illuminated}) in the expression of $\psi(\bx,k)$
given by (\ref{quattordici}), and neglecting the second term (which is certainly legitimate
for small values of $v$, i.e., near the boundary between the two media \cite{Ludwig}) we get the following approximation:
\beq
\label{illuminated3}
\psi(\bx,k) \simeq \sqrt{\pi} g_0(\bx) \frac{1}{k^{1/2}} \frac{e^{\rmi\pi/4}}{v^{1/4}}
\left[e^{\rmi k(u-\frac{2}{3}v^{3/2})}
+e^{-\rmi\pi/2}e^{\rmi k(u+\frac{2}{3}v^{3/2})}\right]~~~~~(v \in \R^+).
\eeq
Proceeding in analogous way, we have for $v\in\R^-$:
\beq
\label{shadow}
\Ai(-k^{2/3}v)=\frac{1}{2\sqrt{\pi}}w_-(-k^{2/3}v)~~~~~(v \in \R^-).
\eeq
For sufficiently high frequency we can approximate $w_-$ with $w_-^{(\rm a)}$, obtaining:
\beq
\label{new-shadow}
\Ai(-k^{2/3}v) \simeq
\frac{1}{2\sqrt{\pi}}\frac{1}{k^{1/6}}\frac{e^{-\frac{2}{3}k(-v)^{3/2}}}{(-v)^{1/4}}~~~~~(v \in \R^-).
\eeq
Again, plugging the r.h.s. of (\ref{new-shadow}) in the expression of $\psi(\bx,k)$, given by (\ref{quattordici}),
we have:
\beq
\label{shadow2}
\psi(\bx,k) \simeq \sqrt{\pi} g_0(\bx) \frac{1}{k^{1/2}}
\frac{1}{(-v)^{1/4}} \, e^{\rmi ku} \, e^{-k\frac{2}{3}(-v)^{3/2}} ~~~~~(v \in \R^-).
\eeq
Note that $u+\rmi \frac{2}{3}(-v)^{3/2}$ is the complex--valued phase.

~

\noindent
{\em Remark}. The exponentials which are contained in the functions $w_\pm^{(\rm a)}(-k^{2/3}v)$ are of the type
$\exp(\pm\frac{2}{3}k(-v)^{3/2})$ and, consequently, we have
$\Imag\phi_\pm(v) = \pm\frac{2}{3}k|v|^{3/2}\sin\frac{3}{2}(\ph v+\pi)$.
In this case we see that care must be taken in the application of the
prescription $\Imag[\phi_+(v)-\phi_-(v)]=0$ for the localization of Stokes lines;
in fact, this would lead to identify erroneously three Stokes lines at $\ph v = \pm\frac{\pi}{3}, \pi$. The presence of the
Stokes line at $\ph v=\pi$ would imply the existence of a second ray in the shadow, which actually
does not exist. However, as discussed in the Appendix
and also shown in Fig. \ref{figure_2}b, the line at $\ph v=\pi$ is not a Stokes line,
so that only the wave with subdominant character, given in (\ref{shadow2}),
is indeed present in the shadow.

\section{Generation of evanescent waves at total reflection}
\label{se:generation}
In this section we obtain the formulae which describe the wave propagation when the
wavefronts are represented by straight lines in $\R^2$, focusing on the total reflection at the boundary between the
denser and the rarer medium.
Then we introduce in $\R^2$ the system of orthogonal axes (O;X,Y), and denote by $x$ and $y$ the corresponding
Cartesian coordinates. The interface between the two media is represented by the line $y=0$.
We treat the three regions separately: first, the illuminated region lying in the upper half--plane ($y>0$) with
refractive index $n>1$; then the shadow region ($y<0$) with unitary refractive index; finally, the boundary between
the two media at $y=0$.

Besides the Ludwig eikonal system (\ref{quindici}) here we will consider also the two equations which correspond to the
transport equation (\ref{sette});
following the procedure outlined in Section \ref{se:collision}, that is substituting Eq. (\ref{quattordici}) into
the Helmholtz equation (\ref{uno}) and collecting powers of $k$, we obtain the following equations that govern the
amplitudes $g_0(\bx)$ and $h_0(\bx)$ \cite{Ludwig}:
\begin{subequations}
\label{transport}
\begin{eqnarray}
2\nabla u\cdot\nabla g_0+\Delta u g_0+2v\nabla v\cdot\nabla h_0+v\Delta v h_0+(\nabla v)^2 h_0 &=& 0,\label{transporta} \\
2\nabla v\cdot\nabla g_0+\Delta v g_0+2\nabla u\cdot\nabla h_0+\Delta u h_0 &=& 0. \label{transportb}
\end{eqnarray}
\end{subequations}
Also system (\ref{transport}), which is linear in $g_0$ and $h_0$, features a mixed--type character similar to the
one displayed by the eikonal system (\ref{quindici}): i.e., it is hyperbolic where $v>0$, and elliptic where $v<0$.
This characteristic will be discussed with more details in the next subsections
(the transport problem can also be treated by means of the Leontovich--Fock parabolic equation;
the interested reader is referred to \cite{Babich2,Babich1}).

\subsection{The illuminated region: $\boldsymbol{y>0}$}
\label{subse:illuminated}
We take $u(x)=nx\sin\theta_i$, where
$\theta_i$ is the incidence angle computed, as usual, with respect
to the normal to the interface, i.e., the $Y$--axis, and $n$ is the
refractive index of the denser medium. The critical angle
$\theta_\ell$, at which total reflection sets in, is given by
$\sin\theta_\ell=1/n$. First let us consider the eikonal system
(\ref{quindici}). From the expression of $u(x)$ we have
$u_x=n\sin\theta_i$, and, from Eq. (\ref{quindicib}) it follows
that $v_x=0$. From Eq. (\ref{quindicia}) we have:
\beq
v\left(\frac{\rmd v}{\rmd y}\right)^2=n^2\cos^2\theta_i,
\label{e1}
\eeq
from which it follows that in the illuminated region the sign of
$v$ is always positive. Integrating (\ref{e1}) with the constraint
that $v(x,y)>0$ for $y>0$, we obtain, for $n= {\rm constant}$:
\beq
\smfr{2}{3}v^{3/2} = ny\cos\theta_i ~~~~~~~~~(y>0).
\label{e2}
\eeq
Then the constant phase lines $\Phi^\pm(u,v)=u\pm\frac{2}{3}v^{3/2}=\Cs$ are given, in the $(x,y)$--plane, by
$\Phi^\pm(x,y)=x\sin\theta_i \pm y\cos\theta_i = \Cs$. Substituting the expressions of $u(x,y)$ and
$v(x,y)$ in formula (\ref{illuminated3}), we have:
\beq
\psi_I(x,y,\theta_i;k)\simeq \frac{G_I(x,y,\theta_i)}{\sqrt{k}}
\left[e^{\rmi kn(x\sin\theta_i-y\cos\theta_i)}+e^{-\rmi\pi/2}\,e^{\rmi kn(x\sin\theta_i+y\cos\theta_i)}\right],
\label{e3}
\eeq
where
$G_I(x,y,\theta_i)=\sqrt{\pi}e^{\rmi\pi/4}(2/3n\cos\theta_i)^{1/6}(g_0(x,y)/y^{1/6})$,
and the subscript $I$ stands for recalling that we are considering
the illuminated region. Formula (\ref{e3}) factorizes the
wavefunction $\psi_I$ as the product of the amplitude
$(G_I/\sqrt{k})$ times the exponentials representing the ray
propagation: the incident ray
$\exp[\rmi kn(x\sin\theta_i-y\cos\theta_i)]$ travelling along the direction
$(\sin\theta_i,-\cos\theta_i)$ with wavenumber $kn$, $k$ being the
wavenumber in the vacuum (recall that a time dependence
$e^{-\rmi\omega t}$ is understood and omitted), and the reflected
ray $\exp[\rmi kn(x\sin\theta_i+y\cos\theta_i)]$ travelling along
the direction $(\sin\theta_i,\cos\theta_i)$. At the grazing angle,
i.e., $\theta_i=\pi/2$, the wavefunction is given by
$\psi_I=(\sqrt{2}e^{-\rmi\pi/4}G_I/\sqrt{k})e^{\rmi knx}$,
which represents a ray travelling along the $X$--axis, and no
reflected ray is present, as it should be.

~

\noindent
{\em Remark}. There is a very unpleasant ambiguity in the literature concerning the sign of the phase--shift
$e^{\pm\rmi\pi/2}$ associated with the reflected ray.
This ambiguity is closely connected with an analogous one encountered in
connection with the Maslov indexes \cite{Delos}. In formulae (\ref{e3}) the sign
of the phase--shift (i.e., the minus sign in $e^{-\rmi\pi/2}$) plays a relevant role in order to give a
correct meaning to the physical displacement of the wavefront. Furthermore, this sign follows from our analysis
of the asymptotic behavior of the Airy function (see formula (\ref{A18}) of the Appendix), which, in some sense,
completes the Dingle's representation. Last but not least, this sign is in accord with the following
Keller's prescription, given in a different but contiguous problem: there is a phase--shift of the
form $e^{-\rmi\pi/2}$ along a path in the direction of propagation; the sign must be reversed in the opposite direction
\cite{Keller}.

~

Let us now consider the transport system (\ref{transport}).
From the expression of $u(x,y)$ and $v(x,y)$ we have:
$\Delta u=0$, $v_y=(2n^2\cos^2\theta_i/3y)^{1/3}$,
$v_{yy}=-((2n^2\cos^2\theta_i)^{1/3}/(3y)^{4/3})$,
which substituted into (\ref{transport}) yield the following first--order system of partial differential equations:
\begin{subequations}
\label{transport_illuminated}
\begin{eqnarray}
&& \frac{\partial g_0}{\partial y}+\beta_I \, y^{1/3}\,\frac{\partial h_0}{\partial x}=
\frac{1}{6y}\,g_0, \label{transport_illuminateda} \\
&& \frac{\partial h_0}{\partial y}+\alpha_I \, y^{-1/3}\,\frac{\partial g_0}{\partial x} =
-\frac{1}{6y}h_0, \label{transport_illuminatedb}
\end{eqnarray}
\end{subequations}
where $\alpha_I=\tan\theta_i(\frac{3}{2}n\cos\theta_i)^{-1/3}$ and
$\beta_I=\tan\theta_i(\frac{3}{2}n\cos\theta_i)^{1/3}$.
It is convenient to recast system (\ref{transport_illuminated})
in vector form by introducing the unknown 2-vector (column vector) ${\bf q}(x,y) = (g_0,h_0)^T$; then,
system (\ref{transport_illuminated}) becomes:
\beq
\label{transport_illuminated_matrix}
{\bf q}_y + B {\bf q}_x = C {\bf q},
\eeq
where the $2\times 2$ matrices $B$ and $C$ read
\beq
\label{matrices}
B = \left (
\begin{array}{cc}
0 & \beta_I \, y^{1/3} \\
\alpha_I \, y^{-1/3} & 0
\end{array}
\right),
~~~~~~
C = \frac{1}{6y}\left (
\begin{array}{cc}
1 & 0 \\
0 & -1
\end{array}
\right).
\eeq
The characteristic curves associated with system (\ref{transport_illuminated_matrix}) can be
obtained from the characteristic differential equation \cite{John}
\beq
\label{char-eq}
\frac{\rmd x}{\rmd y} = \lambda_\pm = \pm\sqrt{\alpha_I\beta_I}=\pm\tan\theta_i,
\eeq
where $\lambda_\pm$ denotes the two eigenvalues of the matrix $B$.
The characteristic curves are then given by: $C^\pm(x,y)=x\pm y \tan\theta_i=\Cs$.
Since the characteristics are real and distinct,
system (\ref{transport_illuminated_matrix}) (and also system (\ref{transport_illuminated})) is of
hyperbolic type.
It is important to notice that the amplitude characteristic curves $C^+=\Cs$ are orthogonal to the
constant phase lines
$\Phi^-=\Cs$, while the curves $C^-=\Cs$ are orthogonal to the lines $\Phi^+=\Cs$.
System (\ref{transport_illuminated_matrix}) can be transformed in a more convenient form by introducing
the new unknown column vector ${\bf p}=(p^{(1)},p^{(2)})^T$ by the relation ${\bf q}=\Gamma {\bf p}$, where $\Gamma$
denotes the matrix whose columns are given by the real eigenvectors $\xi_\pm$
of the matrix $B$, corresponding to the eigenvalues $\lambda_\pm$. The matrix $\Gamma$ reads:
\beq
\label{eigenvectors}
\Gamma = \left (
\begin{array}{cc}
\beta_I \, y^{1/3} & \tan\theta_i \\
-\tan\theta_i & \alpha_I \, y^{-1/3}
\end{array}
\right).
\eeq
Therefore, system (\ref{transport_illuminated_matrix}) is transformed in the following couple of differential
equations for the functions $p^{(1)}$ and $p^{(2)}$:
\begin{subequations}
\label{uncoupled-system}
\begin{eqnarray}
&& p^{(1)}_y+\lambda_- p^{(1)}_x = -\frac{1}{6y}p^{(1)} ~~~~~~~(\lambda_-=-\tan\theta_i), \label{uncoupled-systema} \\
&& p^{(2)}_y+\lambda_+ p^{(2)}_x = \frac{1}{6y}p^{(2)} ~~~~~~~(\lambda_+=+\tan\theta_i). \label{uncoupled-systemb}
\end{eqnarray}
\end{subequations}
Let us now consider Eq. (\ref{uncoupled-systema}); recalling that $\lambda_-=\rmd x/\rmd y$
(see Eq. (\ref{char-eq})), we see that, along the characteristic curves $C^+=\Cs$,
Eq. (\ref{uncoupled-systema}) reduces to the following ordinary differential equation:
\beq
\label{ordinary}
\frac{\rmd p^{(1)}}{\rmd y}=\frac{\partial p^{(1)}}{\partial y}+\frac{\rmd x}{\rmd y}\frac{\partial p^{(1)}}{\partial x}
= -\frac{1}{6y}p^{(1)},
\eeq
whose solution is given by:
\beq
p^{(1)}(x(y),y) = C_1\,y^{-1/6}~~~~~{\rm on~the~curves}~C^+(x,y)=\Cs;~C_1=\Cs.
\label{solution-ordinary_1}
\eeq
It follows that the solution to Eq. (\ref{uncoupled-systema}) will be given by Eq. (\ref{solution-ordinary_1}) times
a function depending only on the characteristic curve $C^+=\Cs$.
With analogous arguments we find that the solution to Eq. (\ref{uncoupled-systemb}) on the
characteristic curves $C^-(x,y)=\Cs$ is given by:
\beq
p^{(2)}(x(y),y) = C_2\,y^{1/6}~~~~~{\rm on~the~curves}~C^-(x,y)=\Cs;~C_2=\Cs.
\label{solution-ordinary_2}
\eeq
Then, the general solution to system (\ref{uncoupled-system}) is given by:
\begin{subequations}
\label{solution-uncoupled-system}
\begin{eqnarray}
p^{(1)}(x,y) &=& f^I_1(x+y\tan\theta_i)\,y^{-1/6}, \label{solution-uncoupled-systema} \\
p^{(2)}(x,y) &=& f^I_2(x-y\tan\theta_i)\,y^{1/6}, \label{solution-uncoupled-systemb}
\end{eqnarray}
\end{subequations}
where $f^I_1(\cdot)$ and $f^I_2(\cdot)$ are two sufficiently regular arbitrary functions. \\
We can now return back to the vector ${\bf q}$ and to the functions $g_0$ and $h_0$ by means of the relation
${\bf q}=\Gamma {\bf p}$: we have

\begin{subequations}
\label{solution}
\begin{eqnarray}
g_0(x,y)\!\!\!\!&=&\!\!\!\!\sqrt{\beta_I}\, y^{1/6}\left[\sqrt{\beta_I}\, f^I_1(x+y\tan\theta_i)+\sqrt{\alpha_I}\,
f^I_2(x-y\tan\theta_i)\right]\!\!,
\label{solutiona} \\
h_0(x,y)\!\!\!\!&=&\!\!\!\!\frac{\sqrt{\alpha_I}}{y^{1/6}}\left[-\sqrt{\beta_I}\, f^I_1(x+y\tan\theta_i)+\sqrt{\alpha_I}\,
f^I_2(x-y\tan\theta_i)\right]\!\!,
\label{solutionb}
\end{eqnarray}
\end{subequations}

Finally, by inserting Eq. (\ref{solutiona}) into Eq. (\ref{e3}) we obtain the expression of the wavefunction
in the illuminated region ($y>0$):
\begin{eqnarray}
\psi_I(x,y,\theta_i;k)\!\!\!\! &\simeq& \!\!\!\!\frac{\psi^0_I}{\sqrt{k}}
\left[\sqrt{\beta_I}\,f^I_1(x+y\tan\theta_i)+\sqrt{\alpha_I}\,f^I_2(x-y\tan\theta_i)\right] \nonumber \\
&&\times
\left[e^{\rmi kn(x\sin\theta_i-y\cos\theta_i)}+e^{-\rmi\pi/2}\,e^{\rmi kn(x\sin\theta_i+y\cos\theta_i)}\right],
\label{illu-final}
\end{eqnarray}
with $\psi^0_I=\Cs$, and $f^I_1(\cdot)$ and $f^I_2(\cdot)$ are (at this stage) arbitrary functions.

The field $\psi_I$ is the product of two factor: the first one, which arises from the transport system (\ref{transport}),
describes the distribution of the amplitude on the lines $x\pm y\tan\theta_i$,
while the second one, which arises form the eikonal system (\ref{quindici}), represents the evolution of the constant
phase lines $x\sin\theta_i\pm y\cos\theta_i=\Cs$.
Obviously, in the wave propagation in a dielectric, homogeneous and isotropic medium, like the one being considered here,
the constant phase lines must be orthogonal to the constant amplitude lines; then from (\ref{illu-final}) we are led to
consider
\begin{eqnarray}
\label{illu-final-4}
\psi_I(x,y,\theta_i;k) &\simeq& \frac{\psi_I^0}{\sqrt{k}}
\left[\sqrt{\beta_I}f^I_1(x+y\tan\theta_i)e^{\rmi kn(x\sin\theta_i-y\cos\theta_i)}\right.\nonumber \\
&&+
\left.
e^{-\rmi\pi/2} \sqrt{\alpha_I}f^I_2(x-y\tan\theta_i) e^{\rmi kn(x\sin\theta_i+y\cos\theta_i)}\right].
\end{eqnarray}
Formula (\ref{illu-final-4}) provides the wavefield in the illuminated region as the sum of two terms:
an incident wave propagating along the direction $(\sin\theta_i,-\cos\theta_i)$ whose amplitude on its
linear wavefronts is described by the function $f^I_1$, and a reflected wave propagating in direction
$(\sin\theta_i,\cos\theta_i)$ with amplitude distribution described by $f^I_2$ on its linear wavefronts.

\subsection{The shadow region: $\boldsymbol{y<0}$}
\label{subse:shadow}
In agreement with the expression of $u(x)$ taken in the illuminated region, we now set:
$u(x)=\sin\theta_t x$ (recall that the refractive index in the region $y<0$ is unitary).
But, now, the incident and the refracted rays invert their role; therefore
we can still use the Snell formula but putting $1/n$ in place of $n$: i.e., writing:
$\sin\theta_i/\sin\theta_t=1/n$.
Therefore, we have $\sin\theta_t=n\sin\theta_i$, and, accordingly,
$u(x)=n\sin\theta_i x$. From Eq. (\ref{quindicib}) it follows that $v_x=0$, and from Eq. (\ref{quindicia})
we have:
\beq
v\left(\frac{\rmd v}{\rmd y}\right)^2=1-n^2\sin^2\theta_i=1-\frac{\sin^2\theta_i}{\sin^2\theta_\ell},
\label{s1}
\eeq
from which it follows that in the region $y<0$ the sign of
$v$ is negative for $\theta_i>\theta_\ell$, i.e., when total reflection occurs. In what follows we will
assume $\theta_i>\theta_\ell$. Integrating (\ref{s1}) with the constraint
that $v(x,y)<0$ for $y<0$, we obtain, for $n={\rm constant}$:
\beq
-\smfr{2}{3}(-v)^{3/2} = \sqrt{n^2\sin^2\theta_i-1}~y~~~~~(y<0; \theta_i>\theta_\ell, v<0).
\label{e6}
\eeq
Substituting the values of $u$ and $v$ in formula (\ref{shadow2}), we obtain:
\beq
\psi_S(x,y,\theta_i;k)\simeq \frac{G_S(x,y,\theta_i)}{\sqrt{k}}
e^{\rmi knx\sin\theta_i}\,e^{k\sqrt{n^2\sin^2\theta_i-1}\,y}~~~~~(y<0),
\label{e7}
\eeq
where $G_S(x,y,\theta_i)=\sqrt{\pi}(\frac{3}{2}\sqrt{n^2\sin^2\theta_i-1})^{-1/6}(g_0(x,y)/(-y)^{1/6})$, and
the subscript $S$ stands for recalling that we are studying the propagation in the shadow region.
Formula (\ref{e7}) factorizes the wavefunction $\psi_S$ as a product of an amplitude $(G_S/\sqrt{k})$
times an evanescent wave which propagates along the positive direction of the horizontal axis,
and which vanishes exponentially along the vertical axis.

Let us now consider the system (\ref{transport}). From the expressions of $u(x,y)$ and $v(x,y)$ we obtain:
$\Delta u=0$, $v_y=-[2(n^2\sin^2\theta_i-1)]^{1/3}(3y)^{-1/3}$,
$v_{yy}=[2(n^2\sin^2\theta_i-1)]^{1/3}(3y)^{-4/3}$, which, inserted into (\ref{transport}), give the
following system of partial differential equations for the functions $g_0(x,y)$ and $h_0(x,y)$ ($y<0)$:
\begin{subequations}
\label{transport_shadow}
\begin{eqnarray}
\frac{\partial g_0}{\partial y}+\beta_S \, y^{1/3} \, \frac{\partial h_0}{\partial x} &=&
\frac{1}{6y}\, g_0, \label{transport_shadowa} \\
\frac{\partial h_0}{\partial y}+\alpha_S \, y^{-1/3} \, \frac{\partial g_0}{\partial x} &=&
-\frac{1}{6y}\, h_0, \label{transport_shadowb}
\end{eqnarray}
\end{subequations}
where $\alpha_S=(2/3)^{1/3}(n^2\sin^2\theta_i-1)^{-2/3}n\sin\theta_i$ and
$\beta_S=-(3/2)^{1/3}(n^2\sin^2\theta_i-1)^{-1/3}n\sin\theta_i$.
Although the structure of this system is very similar to that of
system (\ref{transport_illuminated}) obtained in the illuminated region, the two systems
are profoundly different. In fact, the characteristic differential equation for system (\ref{transport_shadow}) reads:
\beq
\label{char_shadow}
\left(\frac{\rmd x}{\rmd y}\right)^2 = \alpha_S\beta_S = -\frac{n^2\sin^2\theta_i}{n^2\sin^2\theta_i-1}
=-\frac{\sin^2\theta_i}{\sin^2\theta_i-\sin^2\theta_\ell} < 0
~~~~~(\theta_i>\theta_\ell),
\eeq
so that system (\ref{transport_shadow}) does not have real characteristic curves: it is elliptic.
Let us introduce the functions $\overline{g}_0(x,y)$ and $\overline{h}_0(x,y)$, defined by:
\begin{subequations}
\label{funct_shadow}
\begin{eqnarray}
\overline{g}_0(x,y) &=& (-y)^{-1/6} g_0(x,y), \label{funct_shadowa} \\
\overline{h}_0(x,y) &=& (-y)^{1/6} h_0(x,y). \label{funct_shadowb}
\end{eqnarray}
\end{subequations}
Substituting Eqs. (\ref{funct_shadow}) into system (\ref{transport_shadow}) and after suitable differentiation,
we obtian the following couple of differential equations:
\beq
\frac{\partial^2\overline{g}_0}{\partial y^2} - \alpha_S\beta_S\frac{\partial^2\overline{g}_0}{\partial x^2} = 0,
~~~~~
\frac{\partial^2\overline{h}_0}{\partial y^2} - \alpha_S\beta_S\frac{\partial^2\overline{h}_0}{\partial x^2} = 0,
\label{diff_shadow}
\eeq
which are Laplace--type equations ($\alpha_S\beta_S <0$), whose general solution can be written as the
sum of two arbitrary analytic functions
of the complex variables $(x+\rmi\sqrt{-\alpha_S\beta_S}\,y)$ and $(x-\rmi\sqrt{-\alpha_S\beta_S}\,y)$, respectively
\cite{Courant}. Thus, from the solutions of equations (\ref{diff_shadow}) and taking into account
Eqs. (\ref{funct_shadow}) and system (\ref{transport_shadow}) we obtain, for $y<0$:
\begin{subequations}
\label{sol_shadow}
\begin{eqnarray}
g_0(x,y)\!\!\!\!&=&\!\!\!\!
(-y)^{1/6}\,\left[f_1^S(x+\rmi\sqrt{-\alpha_S\beta_S}\,y)+f_2^S(x-\rmi\sqrt{-\alpha_S\beta_S}\,y)\right],
\label{sol_shadowa} \\
h_0(x,y)\!\!\!\!&=&\!\!\!\!\rmi\sqrt{-\frac{\alpha_S}{\beta_S}} (-y)^{-1/6} \nonumber \\
&&\times\left[f_1^S(x+\rmi\sqrt{-\alpha_S\beta_S}\,y)-f_2^S(x-\rmi\sqrt{-\alpha_S\beta_S}\,y)\right].
\label{sol_shadowb}
\end{eqnarray}
\end{subequations}
By inserting Eq. (\ref{sol_shadowa}) into Eq. (\ref{e7}) we finally have the expression of the
wavefunction for $y<0$ and $\theta_i>\theta_\ell$:
\begin{eqnarray}
\psi_S(x,y,\theta_i;k) &\simeq& \frac{\psi^0_S}{\sqrt{k}}
\left[f_1^S(x+\rmi\sqrt{-\alpha_S\beta_S}\,y)+f_2^S(x-\rmi\sqrt{-\alpha_S\beta_S}\,y)\right] \nonumber \\
&&\times \,\, e^{\rmi knx\sin\theta_i}\,e^{k\sqrt{n^2\sin^2\theta_i-1}\,y},
\label{field-shadow}
\end{eqnarray}
with $\psi^0_S=\Cs$, and $f_1^S(\cdot)$ and $f_2^S(\cdot)$ being arbitrary analytic functions.

\subsection{The boundary between the two media: $\boldsymbol{y=0}$, and the Goos--H\"anchen effect}
\label{subse:boundary}
At the interface $y=0$ the refractive index is not defined since it presents a discontinuity, and, therefore,
the eikonal and transport systems can be analyzed only in the limit of $y\rightarrow 0^\pm$.
From the expressions of $v(y)$ in the illuminated and shadow regions, we see that $v(y)\rightarrow 0$ for
$y\rightarrow 0$. Then, in the limit sense, the l.h.s. of Eq. (\ref{quindicia}) reduces to
$|\nabla u|^2$, and we reobtain, in a certain sense, an eikonal equation in standard form, except that
at the r.h.s. of (\ref{quindicia}) is not well-defined.

Since the wavefield and its normal derivative must be continuous, the following boundary
conditions at the interface $y=0$ must be satisfied:
\begin{subequations}
\label{boundary-cond}
\begin{eqnarray}
\psi_I(x,y)|_{y=0^+} &=& \psi_S(x,y)|_{y=0^-}, \label{boundary-conda} \\[+5pt]
\left(\frac{\partial\psi_I}{\partial y}\right)_{y=0^+} &=&
\left(\frac{\partial\psi_S}{\partial y}\right)_{y=0^-}, \label{boundary-condb}
\end{eqnarray}
\end{subequations}
along with appropriate radiation/decay conditions at infinity. \\
The continuity constraint (\ref{boundary-conda}) yields the following relationship among the functions
$f^I_1$, $f^I_2$, $f_1^S$ and $f_2^S$:
\beq
\label{continuity}
\psi^0_S \left[f_1^S(x)+f_2^S(x)\right]=
\psi^0_I\left[\sqrt{\beta_I}f^I_1(x)+e^{-\rmi\pi/2}\sqrt{\alpha_I}f^I_2(x)\right].
\eeq
From the boundary condition (\ref{boundary-condb}),
and taking into account the relation (\ref{continuity}),
we obtain the following relation between the functions $f^I_1$ and $f^I_2$:
\beq
\label{cont-der}
f^I_2(x) = e^{\rmi\pi/2}\sqrt{\frac{\beta_I}{\alpha_I}}
\left(
\frac{\cos\theta_i-\rmi\sqrt{\sin^2\theta_i-\sin^2\theta_\ell}}
{\cos\theta_i+\rmi\sqrt{\sin^2\theta_i-\sin^2\theta_\ell}}\right) f^I_1(x).
\eeq
The term in brackets is of the form $z/z^*$, i.e., it is a phase factor; then (\ref{cont-der})
can be written as
\beq
\label{cont-der-2}
f^I_2(x) = e^{\rmi\pi/2}\sqrt{\frac{\beta_I}{\alpha_I}}\,e^{\rmi\delta(\theta_i)} f^I_1(x),
\eeq
where
\beq
\label{phase-delta}
\delta(\theta_i)=-2\arctan\frac{\sqrt{\sin^2\theta_i-\sin^2\theta_\ell}}{\cos\theta_i}=
-2\arctan\sqrt{-\frac{\alpha_I\beta_I}{\alpha_S\beta_S}}.
\eeq
By inserting (\ref{cont-der-2}) into (\ref{illu-final-4}), we obtain
the following expression of the field in the illuminated region $(y>0)$:
\begin{eqnarray}
\label{illu-final-5}
\psi_I(x,y,\theta_i;k) &\simeq & \frac{\psi^0}{\sqrt{k}}
\left[f^I_1(x+y\tan\theta_i)e^{\rmi kn(x\sin\theta_i-y\cos\theta_i)}\right. \nonumber \\
&&+ \left.
e^{\rmi\delta(\theta_i)} f^I_1(x-y\tan\theta_i) e^{\rmi kn(x\sin\theta_i+y\cos\theta_i)}\right],
\end{eqnarray}
$\psi^0 = \Cs$.
Finally, condition (\ref{boundary-condb}) and formula (\ref{cont-der-2}) yield the following expressions
for the functions $f^I_1(x)$, $f^S_1(x)$:
\begin{subequations}
\label{shadow-condition}
\begin{eqnarray}
\psi^0_S\,f_1^S(x)&=&\psi^0_I\sqrt{\beta_I}\,[1+\rmi\tan(\delta/2)]\,f_1^I(x)+C, \label{shadow-conditiona} \\[+3pt]
\psi^0_S\,f_2^S(x)&=&\psi^0_I\sqrt{\beta_I}\,[1+\rmi\tan(\delta/2)]\,\cos\delta\, f_1^I(x)-C, \label{shadow-conditionb}
\end{eqnarray}
\end{subequations}
$C$ being an arbitrary constant. Plugging equations (\ref{shadow-condition}) into (\ref{field-shadow}), we
finally have the expression of the field in the shadow region $(y<0)$:
\begin{eqnarray}
\psi_S(x,y,\theta_i;k) &\simeq& \frac{\psi^0}{\sqrt{k}}\,[1+\rmi\tan(\delta/2)] \nonumber \\
&&\times\left[f_1^I(x+\rmi\sqrt{-\alpha_S\beta_S}\,y)+\cos\delta\,f_1^I(x-\rmi\sqrt{-\alpha_S\beta_S}\,y)\right]
\nonumber \\
&&\times \,\, e^{\rmi knx\sin\theta_i}\,e^{k\sqrt{n^2\sin^2\theta_i-1}\,y},
\label{field-shadow-final}
\end{eqnarray}
The reflection coefficient $e^{\rmi\delta(\theta_i)}$ in (\ref{illu-final-5}) describes the
phase change that the reflected ray undergoes at total reflection, and coincides with
the phase change given by the Maxwell's theory \cite{Born,Puri} for a linearly polarized TE wave
(also called $s$ polarization).
It is worth noting that this phase--shift is intimately related to the mixed
hyperbolic--elliptic character of the Ludwig's transport system (\ref{transport}) since, as shown by
the formula of $\delta(\theta_i)$ in (\ref{phase-delta}), it depends on the ratio
between the characteristic determinants of system (\ref{transport}) in the illuminated and in the shadow regions.

\begin{figure}[tb]
\centering
\includegraphics[width=4.5in]{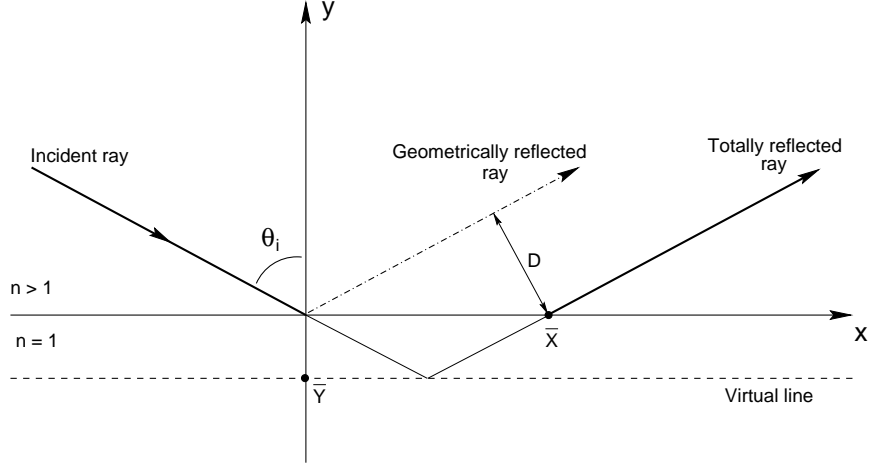}
\caption{\label{figure_goos}\small
Path of a light ray at total reflection. The incident ray is not geometrically reflected at the
interface $y=0$, but at the virtual line located at $y=\overline{Y}$ into the rarer medium, so that it emerges
shifted by $D$ in the denser medium.}
\end{figure}

The role of $e^{\rmi\delta(\theta_i)}$ can be easily illustrated.
The wavefront associated with the reflected ray is delayed by the factor $e^{\rmi\delta}$ (notice that
$\delta\leqslant 0$), so it appears as if it would have been reflected by some virtual surface located a small
distance within the rarer medium and not by the actual physical interface (see Fig. \ref{figure_goos}).
We thus find a behavior already conjectured by Newton \cite{Newton}: at the condition of total internal reflection
at the boundary between an optically denser and an optically rarer medium, a beam of light
penetrates some distance into the rarer medium, and then re--emerges in the optically denser medium.
Consequently, the actual reflected (i.e., totally reflected) ray is laterally shifted with respect to a ray
geometrically reflected (see Fig. \ref{figure_goos}). This type of behavior has been experimentally verified
by Goos and H\"anchen indeed \cite{Goos}, and one can thus speak of Goos--H\"anchen effect \cite{Goos,Lotsch}.
The magnitude of this lateral shift, which is found experimentally to be of the order of a few wavelenghts,
can be evaluated as follows \cite{Artmann,Li1,Li2,McGuirk}.
Since $\delta$ is a function of the incidence angle $\theta_i$, it can be thought of as a function of
the $x$--component $k_x=kn\sin\theta_i$ of the wavenumber vector; then, expanding $\delta(k_x)$ in a
Taylor series about $k_x=0$, and truncating at the linear term, we have:
\beq
\label{g1}
\delta(k_x)=\delta_0+\overline{X}k_x + O(k_x^2),~~~~~
\overline{X}=\frac{\lambda}{2\pi n}\frac{1}{\cos\theta_i}\frac{d\delta}{d\theta_i} \leqslant 0,
\eeq
where $\lambda$ is free--space wavelength.
Then, substituting (\ref{g1}) into formula (\ref{illu-final-5}) we have
\begin{eqnarray}
\label{g2}
\psi_I(x,y,\theta_i;k) &\simeq& \frac{\psi^0}{\sqrt{k}}
\left[f^I_1(x+y\tan\theta_i)e^{\rmi kn(x\sin\theta_i-y\cos\theta_i)}\right. \nonumber \\
&&+ \left.
e^{\rmi\delta_0} f^I_1(x-y\tan\theta_i) e^{\rmi kn[(x+\overline{X})\sin\theta_i+y\cos\theta_i]}\right], ~~~~~
\end{eqnarray}
that is, the reflected ray appears shifted by $|\overline{X}|$ in the positive $x$ direction (recall that
$\overline{X}<0$). From the geometry shown in Fig. \ref{figure_goos}, the measured Goos--H\"anchen lateral
ray--shift $D$ can be easily computed; we have:
\beq
\label{g3}
D=\overline{X}\cos\theta_i=\frac{\lambda}{2\pi n}\frac{d\delta}{d\theta_i},
\eeq
which is identical to the classical expression obtained by Artmann in \cite{Artmann} (see also \cite{McGuirk}).

Finally, by inspection of (\ref{illu-final}) and in view of formula (\ref{cont-der-2}) relating
$f^I_1(x)$ and $f^I_2(x)$, we can see that, if the function $f^I_1$ is constant,
then the factor in the first square brackets in (\ref{illu-final}) becomes merely a complex constant,
and the phase--shift between the incident and the reflected ray reduces to $e^{-\rmi\pi/2}$: i.e.,
$\delta(\theta_i)$ does not play any role.
Then the Goos--H\"anchen shift due to $\delta(\theta_i)$ appears only when the amplitude on the wavefront
is not constant, in particular
when the incident wave is a light beam with limited transverse extent
(see also \cite{Horowitz,Lotsch,McGuirk,Sommerfeld} for a discussion of the Goos--H\"anchen effect
in the case of incident beams with finite transverse width).

\appendix
\section*{Appendix: Borel summation and the Stokes switching mechanism}
\label{appendix}
The main purpose of this appendix is to show how to locate Stokes and anti--Stokes lines
without invoking any physical, though reasonable, argument, but working out the problem exclusively by the use of
mathematical tools. We split this appendix into three parts. The first part is devoted to the Borel summation
which, after Dingle \cite{Dingle}, is the main mathematical tool used for investigating the Stokes phenomenon.
The second and the third parts are devoted to the discussion of the Stokes and anti--Stokes lines, respectively.

\noindent
{\bf (a) Borel summation.} Consider the formal series: $\sum_{n=0}^\infty (-1)^n (n!/x^n)$.
Following the Borel method, we formally write:
\beq
\sum_{n=0}^\infty (-1)^n \frac{n!}{x^n}=
\int_0^{+\infty}e^{-t}\left[\sum_{n=0}^\infty\left(\frac{-t}{x}\right)^n\right]\,\rmd t.
\label{A1}
\eeq
We can give a meaning to the integral at the r.h.s. of (\ref{A1}) in view of the following considerations:
the series $\sum_{n=0}^\infty(-t/x)^n$ converges, in the domain $|t|<x$ $(x>0)$, to a function
$1/(1+(t/x))$, which can be continued analytically outside the region $|t|<x$ to the whole semiaxis
$[0,+\infty)$, and which is dominated by the exponential $e^{-t}$. Therefore the integral
\beq
\Lambda_0(x) = \int_0^{+\infty} \frac{e^{-t}}{1+(t/x)}\,\rmd t
\label{A2}
\eeq
converges, and represents the Borel sum of the formal series at the l.h.s. of (\ref{A1}). Next,
we can perform the analytic continuation of $\Lambda_0(x)$ from $\R^+$ to $\C$. The integrand of the function
$\Lambda_0(z)$ presents a pole at $t=-z$, and therefore, when one turn is made around the origin, we get:
\beq
\Lambda_0(ze^{\pm 2\pi\rmi}) = \Lambda_0(z) \mp 2\pi\rmi z e^{z},
\label{A3}
\eeq
where the factor $e^{\pm 2\pi\rmi}$ is for indicating that one turn is made around the
origin in counterclockwise ($+$) or clockwise ($-$) sense, respectively.
Proceeding in an analogous way, we can perform a summation, in the sense of Borel, not of the whole
series but only of the late terms. We have:
\beq
\sum_{m=n}^\infty (-1)^m \frac{m!}{x^m}=
\int_0^{+\infty}e^{-t}\left[\sum_{m=n}^\infty\left(\frac{-t}{x}\right)^m\right]\,\rmd t=
\frac{1}{(-x)^n}\int_0^{+\infty}\frac{t^n\,e^{-t}}{1+(t/x)}\,\rmd t.
\label{A4}
\eeq
We are thus led to write:
\beq
\sum_{m=n}^\infty (-1)^m \frac{m!}{x^m}=\frac{n!}{(-x)^n}\Lambda_n(x),
\label{A5}
\eeq
where $\Lambda_n(x)$ is given by
\beq
\Lambda_n(x) = \frac{1}{n!} \int_0^{+\infty} \frac{t^n\,e^{-t}}{1+(t/x)}\,\rmd t.
\label{A6}
\eeq
Next we obtain a formula, analogous to (\ref{A3}), of the following form:
\beq
\Lambda_n(ze^{\pm 2\pi\rmi}) = \Lambda_n(z) \mp 2\pi\rmi \frac{e^{\pm\rmi\pi n} z^{(n+1)} e^{z}}{n!}.
\label{A7}
\eeq

\noindent
{\bf (b) Stokes switching mechanism.} Let us come back to the Airy equation (\ref{ventinove});
applying the WKB ansatz, that is assuming that $w(z)$ is of the type:
\beq
w(z)=h(z)\,\exp\left(\pm\int_0^z \zeta^{1/2}\,d\zeta\right),
\label{A8}
\eeq
we obtain the following approximations:
\beq
w_\pm^{(\rm a)}=z^{-1/4}\,\exp\left(\pm\int_0^z \zeta^{1/2}\,d\zeta\right)=
z^{-1/4}\,\exp\left(\pm\smfr{2}{3}z^{3/2}\right)~~~~(z\in\R^+).
\label{A9}
\eeq
The complete asymptotic expansions of two solutions to the Airy equation are given by:
\beq
w_\pm= z^{-1/4}\,\exp\left(\pm\smfr{2}{3}z^{3/2}\right)\,W_\pm(z)~~~~(z\in\R^+),
\label{A10}
\eeq
where the corrective terms $W_\pm(z)$ are represented by the formal asymptotic series
$\sum_0^\infty(\pm 1)^m W_m$. Using standard notations
we have:
\beq
\Ai(z) = \frac{1}{2\sqrt{\pi}}\,w_-(z)~~~~~\Bi(z) = \frac{1}{\sqrt{\pi}}\,w_+(z)~~~~~(\ph z = 0),
\label{A11}
\eeq
and we now focus on $\Ai(z)$.
Then we have (see \cite{Dingle}):
\beq
W_-(z)=\sum_{m=0}^\infty (-1)^m W_m(z) =
\sum_{m=0}^{n-1} (-1)^m W_m(z) + \sum_{m=n}^\infty (-1)^m W_m(z).
\label{A12}
\eeq
Dingle \cite{Dingle} has been able to derive the expression of the late terms $W_m(z)$ in the sum
at the r.h.s. of (\ref{A12}). One has:
\beq
W_m(z) = \frac{1}{2\pi[\cF(z)]^m} \sum_{s=0} [-\cF(z)]^s\,(m-s-1)!\,W_s(z) ~~~~~(m \gg 1),
\label{A13}
\eeq
where $\cF(z)=\frac{4}{3}z^{3/2}$. Substituting (\ref{A13}) in $\sum_{m=n}(-1)^m W_m$, and using
formula (\ref{A5}), we obtain:
\beq
\sum_{m=n}^\infty(-1)^m W_m=
\frac{1}{2\pi(-\cF)^n}\sum_{s=0}(-\cF)^s (n-s-1)!\,W_s\,\Lambda_{n-s-1}(\cF)~~~~~(|\ph \cF|<\pi).
\label{A14}
\eeq
Let us now explore if there exist and where the Stokes lines associated with
$W_-(z)$ are located (see Fig. \ref{figure_2}a).
Recalling that the integrand in $\Lambda_n(z)$ presents a pole at $t=-z$, which, according to formula (\ref{A7}),
generates a discontinuity
of the form $\mp 2\rmi\pi(e^{\pm\rmi\pi n}z^{(n+1)}e^z)/n!$, we can say that the unique Stokes line,
associated with $W_-(z)$, lies at $\ph\cF=\pi$
(see formula (\ref{A14})), i.e., at $\ph z = \frac{2}{3}\pi$, recalling that $\cF=\frac{4}{3}z^{3/2}$.
Next we apply directly formula (\ref{A7}) to $\Lambda_{n-s-1}(\cF)$; this amounts to take
into account a term of the form
$2\rmi\pi(-\cF)^{(n-s)}e^{\cF}/(n-s-1)!$. Therefore, the asymptotic expansion of $\Ai(z)$, which is given by
the first equality in (\ref{A11}) when $\ph z$ varies in the range $0\leqslant\ph z < \frac{2}{3}\pi$,
changes, crossing the Stokes line at $\ph z = \frac{2}{3}\pi$, and we have:
\begin{eqnarray}
\Ai(z)&=&\frac{1}{2\sqrt{\pi}}z^{-1/4}\left[
e^{-\cF/2}\sum_{m=0}^\infty(-1)^m W_m+\rmi e^{\cF/2}\sum_{s=0}^\infty W_s\right] \nonumber \\
&&= \frac{1}{2\sqrt{\pi}}\left(w_-+\rmi w_+\right).
\label{A15}
\end{eqnarray}
At this point one can easily verify two of the Dingle's rules \cite{Dingle}:
\begin{itemize}
\item[(i)] at $\ph z = \frac{2}{3}\pi$ (i.e., $\ph\cF=\pi$) all the late terms of $W_-(z)$ are homogeneous in phase
and all of the same sign;
\item[(ii)] crossing the Stokes ray at $\ph z = \frac{2}{3}\pi$, $\sum_{m=n}^\infty (-1)^m W_m$ generates a discontinuity,
according to formula (\ref{A7}), which is, on the ray, $\frac{\pi}{2}$ out of phase with the series, and
proportional to its associated function.
\end{itemize}

Representation (\ref{A15}) contains the function $w_+$, and, therefore the asymptotic series $W_+$. Then,
in order to obtain its range of validity, we have to find the Stokes lines associated with $W_+$. From
the definition of $W_+$ and by means of arguments similar to those used above for $W_-$, we have that the Stokes lines
associated with $W_+$ are located at the phases for which $\ph(-\cF)=\pi$, that is $\ph(\cF)\pm\pi=\pi$;
this means that the Stokes lines of $W_+$ are located at $\ph z=0,\frac{4}{3}\pi$.
It follows that representation (\ref{A15}) holds in the range $\frac{2}{3}\pi<\ph z<\frac{4}{3}\pi$.

Finally, since both $W_\pm$ have no Stokes lines in the range $\frac{4}{3}\pi<\ph z<2\pi$, the representation
of $\Ai(z)$ in (\ref{A11}) can be extended to this phase range.

Summarizing, we have:
\beq
\label{A18}
\renewcommand{\arraystretch}{1.2}
\Ai(z)=\frac{1}{2\sqrt{\pi}}\,\times \left\{
\begin{array}{ll}
w_-(z) & ~~(0\leqslant\ph z < \smfr{2}{3}\pi)\cup(\smfr{4}{3}\pi < \ph z < 2\pi), \\
w_-(z) + \rmi w_+(z) &~~ \smfr{2}{3}\pi < \ph z < \smfr{4}{3}\pi.
\end{array}
\right .
\eeq
It is worth remarking that the Stokes lines are associated with the asymptotic expansion of a particular solution
to the differential equation (\ref{ventinove}) and not with the differential equation itself. For instance,
to the expansion $w_-(z)$ is associated the Stokes line at $\ph z = \frac{2}{3}\pi$, while the lines at
$\ph z = 0, \frac{4}{3}\pi$ are connected to the expansion $w_+(z)$.
Then, although the asymptotic expansion of $\Ai(z)$ is in general a linear combination of
$w_+(z)$ and $w_-(z)$, $\Ai(z)$ does not have all the Stokes lines associated with $w_\pm(z)$;
in fact, the Stokes line at $\ph z = 0$, which pertains to $w_+(z)$, does not play any role
in the uniform asymptotic expansion of $\Ai(z)$ (see Fig. \ref{figure_2}a).

\noindent
{\bf (c) Anti--Stokes lines.} The modulus of $w_\pm(z)$ is proportional to the function
$\exp\left[\pm\cos\left(\frac{3}{2}\ph z\right)\right]$. The anti--Stokes lines are the lines delimiting
the phase range in which one asymptotic series is exponentially dominant on the other, while on the
anti--Stokes lines the two contributions are comparable.
Then the anti--Stokes lines are located
at those phases such that $\cos\left(\frac{3}{2}\ph z\right)=0$, that is $\ph z=\pm\frac{\pi}{3},\pi$.
In the phase range $\ph z\in\left(-\frac{\pi}{3},\frac{\pi}{3}\right)$ the asymptotic expansion of $\Ai(z)$ contains only
the term $w_-(z)$, which has a subdominant character since $\cos\left(\frac{3}{2}\ph z\right)>0$.
In the ranges $\left(\frac{\pi}{3},\frac{2}{3}\pi\right)$ and $\left(-\frac{\pi}{3},-\frac{2}{3}\pi\right)$ the Airy
function $\Ai(z)$ is still represented by the sole term $w_-(z)$ which, however, acquires a dominant character since
$\cos\left(\frac{3}{2}\ph z\right)$ is negative. We can thus infer that crossing the anti--Stokes lines
at $\ph z = \pm\frac{\pi}{3}$, $w_-(z)$ varies from a subdominant to a dominant character. As we have shown
in part (b) of this appendix, at $\ph z=\pm\frac{2}{3}\pi$ the term $w_+(z)$ emerges and $\Ai(z)$ is
represented by the combination $(w_-(z)+\rmi w_+(z))$. The term $w_+(z)$ is proportional to
$\cos\left(\frac{3}{2}\ph z\right)$ and has a subdominant character in the ranges
$\left(\frac{2}{3}\pi,\pi\right)\cup(\left(-\frac{2}{3}\pi,-\pi\right)$ since, when $\ph z$ varies in these intervals,
$\cos\left(\frac{3}{2}\ph z\right)$ is negative. For the same reason, $w_-(z)$ (proportional to
$-\cos\left(\frac{3}{2}\ph z\right)$) is dominant.
Let us only note that at $\ph z=\pi$ we have two contributions of comparable amplitude, whereas at $\ph z=0$
we have only one term with subdominant character.

\end{document}